\documentstyle[12pt]{article}
\def\bdot{\dot \beta}
\def\adot{\dot \alpha}
\def\xz{\times}
\def\a{\alpha}
\def\b{\beta}
\def\c{\gamma}\def\C{\Gamma}\def\gdot{\dot\gamma}
\def\d{\delta}
\def\vare{\varepsilon}

\def\m{\mu}

\def\s{\sigma}

\def\th{\theta}

\def\beq{\begin{equation}}\def\eeq{\end{equation}}
\def\beqa{\begin{eqnarray}}\def\eeqa{\end{eqnarray}}
\def\barr{\begin{array}}\def\earr{\end{array}}

\def\tbar{\bar \theta}
\def\th{\theta}
\def\p{\varphi}
\newfont{\bbbold}{msbm10 scaled \magstep1}
\def\com{\mbox{\bbbold C}}

\def\mink{\mbox{\bbbold M}}
\def\flag{\mbox{\bbbold F}}
\def\proj{\mbox{\bbbold P}}
\def\real{\mbox{\bbbold R}}

\def\integer{\mbox{\bbbold Z}}

\newfont{\goth}{eufm10 scaled \magstep1}

\def\gu{\mbox{\goth u}}
\def\gsu{\mbox{\goth su}}
\def\gsl{\mbox{\goth sl}}
\def\uR{\underline{R}}
\begin{document}
\title{$(N,p,q)$ Harmonic Superspace}
\author{G.G. Hartwell and P.S. Howe\\
Department of Mathematics\\
King's College, London}
\maketitle
\begin{abstract}
A family of harmonic superspaces associated with four-dimensional
Minkowski
spacetime is described. Applications are given to free massless
supermultiplets, invariant integrals and super Yang-Mills. The
generalisation
to curved spacetimes is given with emphasis on conformal
supergravities.
\end{abstract}
\vfill\eject
\section{Introduction}
\label{sec:A}
Harmonic superspace was first introduced by Gikos in 1984
\cite{gikos:N=2} and
has proved to be
a useful tool in the analysis of four-dimensional supersymmetric
field
theories, in particular $N=2$ \cite{gikos:N=2} and $N=3$
\cite{gikos:N=3} super
Yang-Mills, the $N=2$ non-linear
sigma model \cite{gikos:sigma1,gikos:sigma2} and $N=2$ supergravity
\cite{gikos:N=2sg,gikos:N=2sg2}.  The harmonic formalism is closely
related
to the projective superspace formalism \cite{klr,lr} and to twistor
theory
\cite{penrose}.

In this article we develop the theory further by studying a family
of
superspaces in four dimensions which includes those previously
studied as
special cases.  We call this family of superspaces $(N,p,q)$
harmonic superspaces (where $p+q \leq N$), and
the subset mentioned above corresponds to $(p,q)=(1,1)$, although
$(3,2,1)$ superspace was also discussed in \cite{gikos:N=3ss}.
We shall introduce
these superspaces firstly in the complex setting where they can be
viewed as
coset spaces of the complex superconformal group $SL(4|N)$; in fact
the
complex superspaces are examples of flag supermanifolds, studied in
particular
by Manin \cite{manin}.  As in twistor theory \cite{wardwells},
one can fit together sets of three
such superspaces into double fibrations and it is in this way that
they are
exploited in, for example, Yang-Mills theory.  In the real setting,
the
double fibration can be replaced by a single fibration over super
Minkowski
space and real $(N,p,q)$ harmonic superspaces have the property that
they
are CR supermanifolds, a CR structure being a generalisation of a
complex
structure, a property observed in the $(1,1)$ case in  \cite{rs}.
Viewed in this light, harmonic analyticity
is related to CR-analyticity.
In a recent paper, an analogous family
of
superspaces for spacetimes of dimension less than four was introduced
\cite{ph:d=3}.
In section \ref{sec:B} we construct the $(N,p,q)$ superspaces and
study
their properties.

{}From the point of view of applications to field theory, it seems
that the
$(p,q)=(1,1)$ superspaces remain the most useful, but we shall show
that
there are field theoretic uses of the more general superspaces.  In
particular, in section \ref{sec:C} we show that the field strength
superfields of some
massless multiplets can be viewed as CR-analytic superfields
on $(N,p,N-p)$ superspaces; in section \ref{sec:D}, we show that
harmonic integration is equivalent to the superaction formulae
introduced in
ref. \cite{ph:actions} and rewrite the 3-loop counter\-term for
linearised
$N=8$
supergravity as a harmonic superspace integral.  This counterterm
was
first constructed by Kallosh \cite{kallosh} in a non-manifestly
symmetric way and
subsequently recast as a superaction in \cite{ph:actions}.  The
version given
here has
the feature of being manifestly supersymmetric.  In section
\ref{sec:E} we
study
super Yang-Mills theory in $(N,p,q)$ harmonic superspaces.
For $(p,q)>(1,1)$, we cannot impose flatness as can
be done for $(p,q)=(1,1)$, but we show that the non-Abelian field
strength
superfield is itself covariantly CR-analytic for appropriate choices
of
$(p,q)$.  Finally, in section \ref{sec:F} we study the conformal
constraints
of
$N$-extended supergravity and show that $(N,p,q)$ superspaces also
have a
r\^ole in this context.
\section{Harmonic Superspaces}
\label{sec:B}
We begin in the complex setting.  The complex superconformal group
is
$SL(4|N)$, the supergroup of $(4+N) \times (4+N)$ complex
supermatrices with
unit superdeterminant, (for $N=4$ this group is not simple).  All
the
superspaces which have been used in supersymmetric field theory can
be
obtained as coset spaces of the form $P \backslash SL(4|N)$ where $P$
is a
parabolic
subsupergroup of $SL(4|N)$.  These spaces are called flag
supermanifolds and
this point of view has been developed at length in an article in
preparation
by the authors \cite{ph:sconf}.  In the present context, the
parabolic
subgroups
have block lower triangular form.  For example, complex $N$-extended
super Minkowski space $\widetilde{\mink}_N$ corresponds to the
subgroup
\begin{equation}
P_N=\left\{ \left( \begin{array}{cccc|ccc}
        \times & \times & & & & & \\
        \times & \times & & & & & \\
\times & \times & \times & \times & \times & . & \times \\
\times & \times & \times & \times & \times & . & \times \\
\hline  \times & \times & & & \times & . & \times \\
        .      & .      & & & .      & . & . \\
        \times & \times & & & \times & . & \times
\end{array} \right) \right\}
\end{equation}
The top left corner corresponds to Minkowski space, while the odd
directions
of the coset correspond to the blank entries in the off-diagonal
blocks.  We
shall use blackboardbold letters to denote complex superspaces, and
tildes to
indicate spaces with compact bodies.

The harmonic superspaces we are interested in have the property that
their
bodies, which correspond to the diagonal part of the subgroup, are
schematically of the form $\widetilde{\mink}_0$ (complex Minkowski
space)
times an internal flag manifold, while they have the same number,
$4N$, of
odd coordinates as super Minkowski space.
We then define {\bf complex (N,p,q) harmonic superspace}
$\widetilde{\mink}_N(p,q)$, $(p+q \leq N)$
as the coset space of $SL(4|N)$ with parabolic subgroup
\begin{equation}
\begin{picture}(300,200)(20,-100)
\put(0,0){
$P_N(p,q)=\left\{ \left( \begin{array}{cccc|ccccccc}
        \times & \times & & & & & & & & & \\
        \times & \times & & & & & & & & & \\
\times & \times & \times & \times & \times & . & . & . & . & . &
\times \\
\times & \times & \times & \times & \times & . & . & . & . & . &
\times \\
\hline  \times & \times & & & \times & . & \times & & & &\\
        .      & .      & & & .      & . & .      & & & &\\
        .      & .      & & & \times & . & \times & & & &\\
        .      & .      & & & \times & . & .      & . & \times & &\\
        .      & .      & & & .      & . & .      & . & .      & &\\
        .      & .      & & & \times & . & .      & . & \times & &\\
        .      & .      & & & \times & . & .  & . & . & . & \times
\\
        .      & .      & & &.& . & .  & . & . & . & . \\
\times & \times & & & \times & . & .  & . & . & . & \times
\end{array} \right) \right\} $ }
\put(275,15)
{$ \left. \phantom{\begin{array}{c} \times \\ .
\\ \times \end{array} } \right\}p $}
\put(275,-72)
{$ \left. \phantom{\begin{array}{c} \times \\ .
\\ \times \end{array} } \right\}q $}
\end{picture}
\end{equation}
The internal space, represented by the bottom right-hand corner, is
the (ordinary) flag manifold $\flag_{p,N-q}(N)$, the space of flags
of type
$(p,N-q)$ in $\com^N$.  (We recall that a flag of type $(k_1 \ldots
k_l)$
in $\com^N$ is a set of subspaces $V_{k_1} \subset V_{k_2} \subset
\ldots
V_{k_l} \subset \com^N$ where $k_1 < k_2 < \ldots < k_l < N$.)  For
each such
superspace we define an associated {\bf (N,p,q) analytic superspace}
$\widetilde{\mink}_{NA}(p,q)$ as the coset space of $SL(4|N)$ with
parabolic
subgroup
\begin{equation}
\begin{picture}(300,200)(20,-100)
\put(0,0){
$P_{NA}(p,q)=\left\{ \left( \begin{array}{cccc|ccccccc}
        \times & \times & & & \times & . & \times & & & &\\
        \times & \times & & & \times & . & \times & & & &\\
\times & \times & \times & \times & \times & . & . & . & . & . &
\times \\
\times & \times & \times & \times & \times & . & . & . & . & . &
\times \\
\hline  \times & \times & & & \times & . & \times & & & &\\
        .      & .      & & & .      & . & .      & & & &\\
        .      & .      & & & \times & . & \times & & & &\\
        .      & .      & & & \times & . & .      & . & \times & &\\
        .      & .      & & & .      & . & .      & . & .      & &\\
        .      & .      & & & \times & . & .      & . & \times & &\\
\times & \times & \times & \times & \times & . & .   & . & . & . &
\times \\
        .      & .      & . & . & .      & . & .      & . & . & . & .
\\
\times & \times & \times & \times & \times & . & .      & . & . & . &
\times
\end{array} \right) \right\}
$ }
\put(280,15)
{$ \left. \phantom{\begin{array}{c} \times \\ .
\\ \times \end{array} } \right\}p $}
\put(280,-72)
{$ \left. \phantom{\begin{array}{c} \times \\ .
\\ \times \end{array} } \right\}q $}
\end{picture}
\end{equation}
Given any simple complex Lie (super)group $G$ and parabolic
sub(super)
groups $P_1, P_2,P_{12}=P_1 \cap P_2$, there is a double fibration
\begin{equation}
\begin{picture}(300,120)(-80,-10)
\put(53,80){$P_{12}\backslash G$}
\put(60,68){\vector(-1,-1){50}}
\put(80,68){\vector(1,-1){50}}
\put(-10,0){$P_2 \backslash G$}
\put(60,0){$\Longleftrightarrow$}
\put(130,0){$P_1 \backslash G$}
\put(11,50){$\pi_L$}
\put(111,50){$\pi_R$}
\end{picture}
\end{equation}
A point $p_1 \in P_1 \backslash G$ corresponds to a subset $\pi_L
\circ
\pi_R^{-1}
(p_1)$ of $P_2 \backslash G$, and a point $p_2 \in P_2 \backslash G$
to a
subset
$\pi_R \circ \pi_L{}^{-1}(p_2)$ of $P_1 \backslash G$.  In the
twistor theory
context
the double fibration allows one to code information about theories in
the
space of interest, $P_1 \backslash G$, into information on the
twistor space
$P_2 \backslash G$ via
the correspondence space $P_{12} \backslash G$.

For each choice of $(N,p,q)$, we have the following double fibration
\begin{equation}
\begin{picture}(300,120)(-80,-10)
\put(49,80){$\widetilde{\mink}_N(p,q)$}
\put(60,68){\vector(-1,-1){50}}
\put(80,68){\vector(1,-1){50}}
\put(-10,0){$\widetilde{\mink}_{NA}(p,q)$}
\put(60,0){$\Longleftrightarrow$}
\put(130,0){$\widetilde{\mink}_N$}
\put(11,50){$\pi_L$}
\put(111,50){$\pi_R$}
\end{picture}
\label{eq:mdf}
\end{equation}
In the case $(p,q)=(1,1)$ it is this mathematical fact
which underlies the interpretation of
the constraints of super Yang Mills theory in terms of analytic
superspace.
In the applications, we are not interested in the compactified
superspaces,
and we define an open set in $\widetilde{\mink}_N$ to be complex
super Minkowski space, $\mink_N$.  We can do this as follows: let
$z \in \mink_N \; (=\com^{4|4N})$ and consider the element $s(z)$ in
$SL(4|N)$ defined by
\begin{equation}
s(z)=\left( \begin{array}{cc|c}
1 & -iX^{\alpha \bdot} & -i \th^{\alpha j} \\ 0 & 1 & 0 \\ \hline
0 & -i \varphi_i{}^{\bdot} & 1
\end{array} \right)
\end{equation}
where $X^{\alpha \bdot}=x^{\alpha \bdot} - \frac{i}{2} \th^{\alpha
i}
\varphi_i{}^{\adot}$.
We can regard $s(z)$ as a section of $SL(4|N)$ considered as a bundle
over
$P_N \backslash SL(4|N)$.  Using standard homogeneous space
techniques one can
then
find how infinitesimal $SL(4|N)$ transformations act on $z$ and
thereby
confirm that $SL(4|N)$ is indeed the complex superconformal group.
The
Maurer-Cartan form on $SL(4|N)$ pulled back to $\mink_N$ is
\begin{equation}
ds(z)s^{-1}(z)=\left( \begin{array}{cc|c}
0 & -iE^{\alpha \bdot} & -iE^{\alpha j} \\ 0 & 0 & 0 \\ \hline
0 & -iE_i{}^{\bdot} & 0
\end{array} \right)
\end{equation}
from which we identify the usual basis $\{E^{\alpha \adot}, E^{\alpha
i},
E_i^{\adot}\}$ of super Minkowski space, namely:
\begin{eqnarray}
E^{\alpha \adot} & = & dx^{\alpha \adot} + \frac{i}{2}d \th^{\alpha
i}
\varphi_i{}^{\adot} + \frac{i}{2} d \varphi_i{}^{\adot} \th^{\alpha
i} \\
E^{\alpha i} & = & d \th^{\alpha i} \\
E_i{}^{\adot} & = & d \varphi_i{}^{\adot}
\end{eqnarray}
We define non-compact harmonic superspace $\mink_N(p,q)$ as
$\pi_R^{-1}(\mink_N)$ and the corresponding analytic superspace
$\mink_{NA}(p,q)$ as $\pi_L \left( \mink_N(p,q) \right)$.
$\mink_N(p,q)$ is in fact simply $\mink_N \times \flag_{p,N-q}(N)$,
so that the

non-compact version of the double fibration (\ref{eq:mdf}) is
\begin{equation}
\begin{picture}(300,120)(-80,-10)
\put(0,80){$\mink_N(p,q)=\mink_N \times \flag_{p,N-q}(N)$}
\put(60,68){\vector(-1,-1){50}}
\put(80,68){\vector(1,-1){50}}
\put(-22,0){$\mink_{NA}(p,q)$}
\put(60,0){$\Longleftrightarrow$}
\put(130,0){$\mink_N$}
\put(11,50){$\pi_L$}
\put(111,50){$\pi_R$}
\end{picture}
\end{equation}
In the following we shall denote the internal flag manifold as simply
$\flag$.
Finally, we need to consider real super Minkowski space, $M_N$, which
can be
defined to be the subspace of $\mink_N$ such that $x$ is real and
$\p_i^{\adot} = \bar{\th}_i^{\adot}$.  Correspondingly, real
$(N,p,q)$
harmonic superspace, $M_N(p,q)$ is $\pi_R^{-1}(M_N) = M_N \times
\flag$.  However, we now find that $\pi_L\left(M_N(p,q)\right)$
embeds $M_N(p,q)$ as a real subsupermanifold of $\mink_{NA}(p,q)$,
the latter
being essentially complex.  Indeed, the compactified versions of
$M_N$ and
$M_N(p,q)$ are coset spaces of the real superconformal group
$SU(2,2|N)$,
whereas there is no corresponding analogue of $\mink_{NA}(p,q)$.  For
this
reason, it is sometimes simpler, in the real context, to consider the
single
fibration
$M_N(p,q) \rightarrow M_N$.  The r\^ole of the analytic space is
then
replaced by CR-analyticity on $M_N(p,q)$ which we now explain.

Let $M$ be a real (super) manifold of dimension $2n+m$, where $n,m
\in
\integer$
for manifolds, and $n,m \in \integer^2$ for supermanifolds.  A {\bf
CR
structure}
on $M$ is a subbundle $K$ of the complexified tangent bundle, $T_c$
of rank
$n$ such that
\begin{eqnarray*}
K \cap \bar{K} & = & \emptyset \\
\left[ K, K \right] & \subset & K
\end{eqnarray*}
where the latter statement means that the commutator of any two
vector fields
belonging to $K$ also belongs to $K$, i.e. $K$ is involutive.  This
notion
generalises that of a complex structure which is recovered in the
case $m=0$.
Given a function $f$ on $M$ we can define a CR operator
$\bar{\partial}_K$ by
\begin{equation}
\bar{\partial}_K f = \pi \circ df
\end{equation}
where $\pi$ is the projection: $1$-forms on $M \rightarrow$ sections
of
$\bar{K}^*$, the dual space of $\bar{K}$.   A function $f$ such that
$\bar{\partial}_K f = 0$ is called CR-analytic, and the involutivity
of $K$
ensures that this is consistent. The holomorphic bundle of vectors
tangent to
the fibre, $T_{\flag}$, is a subbundle of $K$, and it is convenient
to write
\begin{equation}
K=T_{\flag} \oplus T_G
\end{equation}
where $T_G$ is the odd part of $K$. The derivative
$\bar{\partial}_K$ can correspondingly be written as
\begin{equation}
\bar\partial_K=\bar\partial +\bar D
\end{equation}
where $\bar\partial$ is essentially the usual
$\bar{\partial}$
operator,
on $\flag$ and $\bar D$
is the odd part of $\bar\partial_K$. Following Gikos we define a
field $f$ to
be Grassmann analytic, or simply G-analytic, if
\begin{equation}
\bar D f=0
\end{equation}
A G-analytic superfield on $M_N(p,q)$ can be thought of as the
pull-back of a
function defined on $\mink_{NA}(p,q)$ to $M_N(p,q)$.

We shall now describe the above in local coordinates.
It is convenient to work on the space $M_N \times SU(N)$,
as advocated in \cite{gikos:N=2}, which we call
$\hat{M}_N$, instead of on
$M_N(p,q)=M_N \times \flag$ directly.  Since $\flag$ is a coset space
of
$SU(N)$:
\begin{equation}
\flag=S\left(U(p) \times U(p) \times U(N-p-q) \right) \backslash
SU(N)
\end{equation}
a field on $\flag$ is the same as a field on $SU(N)$ invariant under
the
subgroup $P_N(p,q)=S\left(U(p) \times U(p) \times U(N-p-q) \right)$.
More
generally, we can consider fields on $SU(N)$ which take their values
in a
vector space, $V$, which is a
representation space of $P_N(p,q)$ and which are equivariant under
the
action of $P_N(p,q)$, i.e. fields $\phi(u)$, $u \in SU(N)$, such that

\begin{equation}
\phi(hu) = T(h) \phi(u)
\end{equation}
where T(h) is the representation of $P_N(p,q)$ on the vector space
$V$.  Such
fields correspond to sections of various bundles over $\flag$.  In
practice,
they are fields with various $H_N$ indices.  The extension to fields
which
depend also on the coordinates of superspace is straightforward.

Explicitly, we let $u_I{}^i$ be an element of $SU(N)$ where $SU(N)$
acts on
$i$ to the right and $P_N(p,q)$ on $I$ to the left.  We denote the
inverse of
$u$ by $u_i{}^I$.  The right invariant vector fields on $SU(N)$ are
\begin{equation}
D_I{}^J = u_I{}^i \frac{\partial}{\partial u_J{}^i} - \frac{1}{N}
\delta_I{}^J
u_K{}^i \frac{\partial}{\partial u_K{}^i}
\end{equation}
so that
\begin{equation}
D_I{}^J u_K{}^i = \delta_K{}^J u_I{}^i - \frac{1}{N} \delta_I{}^J
u_K{}^i
\end{equation}
Clearly,
\begin{equation}
\left[ D_I{}^J,D_K{}^L \right] = \delta_K{}^J D_I{}^L - \delta_I{}^L
D_K{}^J
\end{equation}
The coordinates of $\hat{M}_N$ are then
$(x^{\alpha \adot}, \th^{\alpha i}, \bar{\th}_i^{\adot}, u_I{}^i)$.
The
supercovariant derivatives dual to $E^{\alpha i}$ and $E_i{}^{\adot}$
are
\begin{eqnarray}
D_{\alpha i} & = & \frac{\partial}{\partial \th^{\alpha i}} -
\frac{i}{2}
\bar{\th}_i^{\adot} \frac{\partial}{\partial x^{\alpha \adot}} \\
\bar{D}_{\adot}^i & = &
-\frac{\partial}{\partial \bar{\theta}_i^{\adot}} + \frac{i}{2}
\th^{\alpha i}
\frac{\partial}{\partial x^{\alpha \adot}}
\end{eqnarray}
and we define
\begin{eqnarray}
D_{\alpha I}= u_I{}^i D_{\alpha i} & &
\bar{D}_{\adot}^I = u_i{}^I \bar{D}_{\adot}^i \nonumber \\
\th^{\alpha I} = u_i{}^I \th^{\alpha i} \; & &
\; \bar{\th}_I^{\adot}=u_I{}^i \bar{\th}_i^{\adot}
\end{eqnarray}
For $(N,p,q)$ superspace, we split the index $I$ up as follows:
\begin{eqnarray}
& I=(R,R'',R') & \nonumber \\
R=1,\ldots,p; \: \: \: \: & R''=1,\ldots,N-(p+q); & \: \: \: \:
R'=1,\ldots,q
\end{eqnarray}
The derivatives corresponding to the isotropy group $P_N(p,q)$ are
\begin{equation}
\left\{ D_R{}^S, D_{R''}{}^{S''}, D_{R'}{}^{S'} \right\}
\end{equation}
and the components of the CR operator $\bar{\partial}_K$ are
\begin{equation}
\{D_{\alpha R}, \bar{D}_{\adot}^{R'}, D_{R}{}^{S''}, D_{R}{}^{S'},
D_{R''}{}^{S'} \}
\end{equation}
while the components of $\bar\partial$ and $\bar D$ are
\begin{equation}
\{D_{R}{}^{S''}, D_{R}{}^{S'},
D_{R''}{}^{S'} \}
\end{equation}
and
\begin{equation}
\{D_{\alpha R}, \bar{D}_{\adot}^{R'}\}
\end{equation}
respectively. A CR-analytic field on $M_N(p,q)$ is defined as a field
$f$ on
$\hat{M}_N$ which satisfies $\bar{\partial}_K f=0$ and which is
equivariant with respect to $P_N(p,q)$.  Since $\flag$ is a compact,
complex manifold, such a field is severely constrained as a function
of
the complex coordinates of $\flag$, because it is a holomorphic
section of a complex vector bundle over $\flag$.  However, G-analytic
fields
are not so constrained. In index notation, a G-analytic field
satisfies
\begin{equation}
D_{\alpha R} f = \bar{D}_{\adot}^{R'} f = 0
\end{equation}
as well as being equivariant under $P_N(p,q)$.

In the case that $p=q$,
i.e. $(N,p,p)$ superspace, we can define a transformation on $SU(N)$
\begin{equation}
u \rightarrow ku = u'
\end{equation}
where
\begin{equation}
k = \left( \begin{array}{ccc} 0 & 0 & 1_p \\
0 & 1_{N-2p} & 0 \\ -1_p & 0 & 0 \end{array} \right)
\end{equation}
Although $k$ acts to the left as $SU(N)$ it is easy to see that it
induces a
transformation of the flag manifold $\flag$ because $khk^{-1} \in H$
for any
$h \in H$.

To define a real structure, we combine the above transformation with
complex
conjugation, i.e.
\begin{equation}
u \rightarrow \overline{ku}=\tilde{u}
\end{equation}
Explicitly
\begin{equation}
\begin{array}{lclclcl}
u_R{}^i & \rightarrow & u_i{}^{R'} &\qquad & u_i{}^R  & \rightarrow &
u_{R'}{}^i \\
u_{R'}{}^i & \rightarrow & -u_i{}^{R} &\qquad & u_i{}^{R'}  &
\rightarrow &
-u_R{}^i \\
u_{R''}{}^i & \rightarrow & u_i{}^{R''} &\qquad & u_i{}^{R''}  &
\rightarrow &
u_{R''}{}^i
\end{array}
\end{equation}
For a field $F(x, \th, u)$ we define
\begin{equation}
\widetilde{F}(x, \th, u) = \overline{F(x,\th,ku)}
\end{equation}
Note that this transformation preserves G-analyticity, since
\begin{eqnarray}
\left(\widetilde{D_{\alpha R} F}\right)(x, \th, u) & = &
\bar{D}_{\adot}^{R'}
\left(\widetilde{F}(x, \th, u)\right) \\
\left(\widetilde{\bar{D}_{\adot}^{R'} F}\right)(x, \th, u) & = &
-D_{\alpha R} \left(\widetilde{F}(x, \th, u) \right)
\end{eqnarray}
It also preserves CR-analyticity, as can easily be checked.  A
general field
$F$ will
transform under a representation of $H_N$.  Depending on the
representation
it may be possible to have real fields, i.e. fields for which
$\widetilde{F}(x, \th, u) = F(x, \th, u)$.

We conclude this section with a brief discussion of the
superconformal group.
Since (compactified) $M_N(p,q)$ is a coset space of $SU(2,2|N)$ we
can easily
find how this
group acts.  However, we are interested in representations of the
Lie
superalgebra ${\mbox{\goth su}}(2,2|N)$ on G-analytic fields,
and it turns out that the
straightforward representation is not the required one.  Instead, we
can show
that the following vector fields, ${\cal V}$, give a representation
of
${\mbox{\goth su}}(2,2|N)$.
\beqa
{\cal V}&=&F^{\a\adot}\partial_{\a\adot}+f^{\a R'}D_{\a R'}+f^{\a
R''}D_{\a
R''}
-\bar f^{\adot}_R\bar D^R_{\adot} -\bar f^{\adot}_{R''}\bar
D^{R''}_{\adot}\nonumber\\
&\phantom{=}&+f_R{}^{S'}D_{S'}{}^R + f_R{}^{S''}D_{S''}{}^R +
f_{R''}{}^{S'}D_{S'}{}^{R''}
\label{eq:V}
\eeqa
where $f^{\a I}=f^{\a i}u_i{}^I,\ \bar f^{\adot}_I=u_I{}^i\bar
f^{\adot}_i$,
and where
\beq
D_{\a i} F^{\b\bdot}+i\d_\a{}^\b \bar{f}_i^{\bdot}=0
\eeq
and
\beq
f_I{}^J= {1\over2}\big( D_{\a I} f^{\a I}-{1\over N}\d_I{}^J D_{\a K}
f^{\a
K}\big)
\eeq
A superconformal vector field on super Minkowski space is given by
$F^{\a\adot}\partial_{\a\adot}+f^{\a i}D_{\a i} - \bar f^{\adot}_i
\bar
D_{\adot}^i$ (for a discussion of the $N=1$ case see, for example,
\cite{cw})
with the
components constrained as above, and the rest of the conditions show
that the
components of ${\cal V}$ are also given in terms of $F$. The vector
field
${\cal V}$ preserves the CR-structure (up to istotropy group terms)
and
G-analyticity. Thus for a scalar $G$-analytic superfield, $f$, we can
define a

superconformal transformation by
\beq
\d f={\cal V}f
\eeq
Note that this is a representation of ${\mbox{\goth su}}(2,2|N)$ on
fields;
it is not the standard representation on fields on $M_N(p,q)$ that
one
constructs using
homogeneous space techniques.

\section{Massless Field Strength Superfields}
\label{sec:C}
It is well-known that a free, massless
supermultiplet of highest spin $s$ can be described
by a field-strength superfield, $W$, which is subject to
certain constraints.  For $N$-extended supersymmetry, when the
greatest spin,
$s<\frac{N}{2}$, $W$ has $2s$ internal indices, is totally
antisymmetric
and satisfies
\begin{eqnarray}
D_{\adot}{}^i
W_{j_1 \ldots j_{2s}} & = & \frac{2(-1)^{2s-1}s}{N-2s+1}
\delta_{[j_1}^i
D_{\adot}^k
W_{j_2 \ldots j_{2s}] k} \nonumber \\
D_{\alpha i}W_{j_1 \ldots j_{2s}} & = &  D_{\alpha [ i} W_{j_1 \ldots
j_{2s}]}
\label{eq:DW}
\end{eqnarray}
(See, for instance \cite{ph:currents}.)  When $s=\frac{1}{4}N$,
there is an additional self-duality condition:
\begin{equation}
\bar{W}^{i_1 \ldots i_{2s}} = \frac{1}{(2s)!}
\varepsilon^{i_1 \ldots i_{2s} j_1 \ldots j_{2s}} W_{j_1 \ldots
j_{2s}}
\end{equation}

These superfields can be described naturally in $(N,p,N-p)$ harmonic

superspace, with $p=2s$, as we shall now demonstrate. It is
convenient to
modify our notation a little by noticing that the isotropy subalgebra
of the
flag manifold $\flag$ is
$\gsu(p)\oplus\gsu(q)\oplus\gsu(r)\oplus\gu(1)\oplus\gu(1)$,
where $r=N-(p+q)$. We can therefore write
\beqa
D_R{}^S &=& \tilde D_R{}^S +\frac{1}{N}\d_R{}^S D_o \\
D_{R'}{}^{S'} &=& \tilde D_{R'}{}^{S'} +\frac{1}{N}\d_{R'}{}^{S'}
D'_o  \\
D_{R''}{}^{S''} &=& \tilde D_{R''}{}^{S''}
+\frac{1}{N}\d_{R''}{}^{S''} D''_o
\eeqa
where
\beq
D_o''=-\frac{pD_o+qD'_o}{r}
\eeq
and where the tilded derivatives correspond to the basis elements of
the
$\gsu$ algebras, so that, e.g. $\tilde D_R{}^R=0$. The $U(1)$ charges
of
the $u$'s, with respect to $(D_o,D'_o)$ are given by
\beq
u_R{}^i:(\frac{(N-p)}{p},-1),\qquad u_{R'}{}^i:
(-1,\frac{(N-q)}{q}),
\qquad u_{R''}{}^i:(-1,-1)
\eeq
with the inverse matrix elements having the opposite charges. For
$(p,q)=(p,N-p)$, the flag manifold $\flag$ is simply the Grassmannian
of
$p$-planes in $\com^N$, and the isotropy algebra is
$\gsu(p)\oplus\gsu(q)\oplus \gu(1)$ (with $q=N-p$). In this
case we have
\beqa
D_R{}^S &=& \tilde D_R{}^S +\frac{1}{N}\d_R{}^S D_o \\
D_{R'}{}^{S'} &=& \tilde D_{R'}{}^{S'} +\frac{1}{N}\d_{R'}{}^{S'}
D'_o
\eeqa
with $D'_o=-(p/q) D_o$. The $U(1)$ charges (with respect to $D_o$) of
$u_R{}^i$
and
$u_{R'}{}^i$ are $q/p$ and $-1$ repectively.

We claim that the on-shell superfield $W_{i_1\dots i_p}$ as defined
above is
equivalent to a CR-analytic superfield, $W$, on $\hat M_N$ which is
invariant
under $\gsu(p)\oplus\gsu(q)$ and has $U(1)$ charge $N-p=q$. First,
suppose we are given $W_{i_1\dots i_p}$. We can define
\beq
W_{R_1\dots R_p}=u_{R_1}{}^{i_1}\dots u_{R_p}{}^{i_p} W_{i_1\dots
i_p}
\eeq
Because of the antisymmetry of $W_{i_1\dots i_p}$ we can set
\beq
W_{R_1\dots R_p}=\vare_{R_1\dots R_p}W
\eeq
where
\beq
W=\frac{1}{p!}\vare^{R_1\dots R_p}u_{R_1}{}^{i_1}\dots
u_{R_p}{}^{i_p}
W_{i_1\dots i_p}
\eeq
To prove that $W$ is G-analytic, first apply $D_{\a R}$,
\beq
D_{\a R} W=\frac{1}{p!}\vare^{R_1\dots R_p}u_R{}^j
u_{R_1}{}^{i_1}\dots
u_{R_p}{}^{i_p}D_{\a j} W_{i_1\dots i_p}
\eeq
Since $D_{\a j} W_{i_1\dots i_p}$ is totally antisymmetric, it
follows that
the product of $u$'s must be also and
hence this product must vanish as $R$ runs only from 1 to $p$. We
also have
\beq
\bar D_{\adot}^{R'}W=\frac{1}{p!}\vare^{R_1\dots
R_p}u_{R_1}{}^{i_1}\dots
u_{R_p}{}^{i_p}u_j{}^{R'}\bar D_{\adot}^j W_{i_1\dots i_p}
\eeq
Since $\bar D_{\adot}^j W_{i_1\dots i_p}$ consists of a sum of terms
with
$\d_i{}^j$, and $u_R{}^i u_i{}^{S'}=0$, it follows that
$\bar D_{\adot}^{R'}W=0$. Hence $W$ is G-analytic. To prove
that it is CR-analytic it is enought to observe that
$D_R{}^{S'}u_T{}^i=0$ which immediately implies that $D_R{}^{S'}
W=0$.

We can also prove the converse, namely, that a CR-analytic field $W$
on
$M_N(p,q)$ with charge $N-p$ defines a superfield $W_{i_1\dots i_p}$
on super
Minkowski space which satisfies the
constraints defining an on-shell superfield with highest
helicity $s=p/2$. To prove this one expands $W$ is harmonics on
$\flag$. Since

$W$ is invariant under $\gsu(p)\oplus\gsu(q)$ it can only depend on
the $u$'s
via the $\gsu(p)\oplus\gsu(q)$-invariants
\beq
a^{i_1\dots i_p}=\frac{1}{p!}\vare^{R_1\dots R_p}u_{R_1}{}^{i_1}\dots
u_{R_p}{}^{i_p}
\eeq
and
\beq
b^{i_1\dots i_q}=\frac{1}{q!}\vare^{R'_1\dots
R'_q}u_{R'_1}{}^{i_1}\dots
u_{R'_q}{}^{i_q}
\eeq
which is actually the dual of the complex conjugate of $a$. Since $a$
has
$U(1)$ charge $q$
and $b$ has $U(1)$ charge $-q$, it follows that $W$ must be of the
form
\beq
W\sim \sum_{k=1} (a)^k (b)^{k-1} W_{(k)}
\eeq
However, $D_R{}^{S'}a=0$ whereas
\beq
D_R{}^{S'}b^{i_1\dots i_q}=\frac{1}{(q-1)!}\vare^{S'R'_2\dots R'_q}
u_R{}^{i_1}
u_{R'_2}{}^{i_2}\dots u_{R'_q}^{i_q}
\eeq
from which it follows that only the $k=1$ term survives in the sum
when we impose $D_R{}^{S'} W=0$. Hence we find
\beq
W=a^{i_1\dots i_p} W_{i_1\dots i_p}
\eeq
where $W_{i_1\dots i_p}$ depends only on $x$, $\th$ and $\tbar$.
Imposing
G-analyticity we easily recover the on-shell constraints described
above as
may easily be checked.

Finally, the self-duality condition on $W_{i_1\dots i_p}$ when
$s=N/4$, i.e.
$N=2p$, is
equivalent to requiring $W$ to be real with repect to the real
structure
introduced in the previous
section. We have
\beq
W=\frac{1}{p!}\vare^{R_1\dots R_p}u_{R_1}{}^{i_1}\dots
u_{R_p}{}^{i_p}
W_{i_1\dots i_p}
\eeq
so
\beq
\tilde W=\frac{1}{p!}\vare_{R'_1\dots R'_p}u_{i_1}{}^{R'_1}\dots
u_{i_p}{}^{R'_p} \bar W^{i_1\dots i_p}
\eeq
Hence, if $\bar W^{i_1\dots i_p}=\frac{1}{p!}\vare^{i_1\dots i_p
j_1\dots j_p}
W_{j_1\dots j_p}$, $W$ is real and vice versa.

\section{Harmonic Integration and Superactions}
\label{sec:D}
Chiral integration in $N=1$ supersymmetry generalises to harmonic
superspace
in a straightforward fashion, as shown by Gikos for the case
$(p,q)=(1,1)$.
In this section we extend this to arbitrary $(p,q)$ and show that
the
resulting integrals are equivalent to a class of superactions of the
type
discussed in \cite{ph:actions}.  As an application, we rewrite the
linearised $N=8$ supergravity three-loop counterterm as an
$(N,p,q)=(8,4,4)$ harmonic superspace integral.

The $G$-analytic measure on $(N,p,q)$ superspace, $d \mu$, is defined
as
follows
\begin{equation}
d \mu = d^4 x \, du \, d^{2r} \th'' d^{2r} \bar{\th}'' \, d^{2p}
\bar{\th} \,
d^{2q} \th'
\end{equation}
where  $\th \sim \th^{\alpha R}, \th' \sim \th^{\alpha R'},
\th'' \sim \th^{\alpha R''}$
and $d u$ is the Haar measure on the coset space $\flag$.  Since
Grassmann
integration
is actually differentiation, one has e.g.
\begin{equation}
d^{2q} \th' \sim (D_{\alpha R'} D_{S'}^{\alpha})^q
\end{equation}
Moreover, since one is integrating over all $\th^{\alpha R}$'s, etc.,
the
measure is invariant under $SU(p) \times SU(q) \times SU(r)$,
$(r=N-(p+q))$,
but it has $U(1)$ charges given by $2(-(N-p+q),(N+p-q))$.  We can
therefore
use
it to integrate $G$-analytic superfields ${\cal L}$ on $\hat{M}_N$
which are
invariant under $SU(p) \times SU(q) \times SU(r)$ and which have
charges
$2(N-p+q,-(N+p-q))$, to get integrals
\begin{equation}
I=\int d \mu \, {\cal L}
\end{equation}
Such integrals are invariant under super Poincar\'{e} transformations
(with
internal symmetry $SU(N)$) and invariant under superconformal
symmetry if
\begin{equation}
\delta {\cal L} = {\cal V} + \Delta {\cal L}
\end{equation}
where ${\cal V}$ is given in (\ref{eq:V}) and
\begin{eqnarray}
\Delta & = & \partial_{\alpha \adot} F^{\alpha \adot} - D_{\alpha
R'}
f^{\alpha R'} - D_{\alpha R''} f^{\alpha R''} +\bar{D}_{\adot}^R
\bar{f}_R^{\adot} + \nonumber \\ & &
\bar{D}_{\adot}^{R''} \bar{f}_{R''}^{\adot}
+D_R{}^{S'} f_{S'}{}^R + D_R{}^{S''} f_{S''}{}^R + D_{R'}{}^{S''}
f_{S''}{}^{R'}
\end{eqnarray}
The simplest example is the case $N=2$, $p=1$, $q=1$, the original
Gikos
example.  In this case we set $u_I{}^i=(u_1{}^i,u_2{}^i)$, where
$u_1{}^i (u_2{}^i)$ has $U(1)$ charge $+1(-1)$ respectively.  The
measure
can be written in the form
\begin{equation}
d \mu = d^4x \, du \, u_1{}^i u_1{}^j u_1{}^k u_1{}^l \, D_{ij}
\bar{D}_{kl}
\end{equation}
where
\begin{equation}
D_{ij} \equiv D_{\alpha i} D_j^{\alpha}, \: \: \: \: \bar{D}^{ij}
\equiv
\bar{D}_{\adot}^i \bar{D}^{\adot j}
\end{equation}
and where internal indices are raised and lowered with the
$\varepsilon$-tensor
for $N=2$.  The integrand ${\cal L}$, which has charge $4$, can be
expanded
in harmonics on $\flag=\com \proj^1=U(1)\backslash SU(2)$ in the
form
\begin{equation}
{\cal L}=u_1{}^i u_1{}^j u_1{}^k u_1{}^l L_{ijkl} + u_1{}^{(i}
u_1{}^j
u_1{}^k u_1{}^l u_1{}^m u_2{}^{n)} L_{ijklmn} + \ldots
\end{equation}
The products of $u$'s in irreducible $SU(2)$ representations are the
spherical harmonics and are orthogonal with respect to the Haar
measure
$du$, so that if we carry out the integration over $\com \proj^1$
only the
first term in the expansion of ${\cal L}$ contributes.  Hence
\begin{equation}
I= \int d \mu {\cal L} = \int d^4 x D^{ij} \bar{D}^{kl} L_{ijkl}
\end{equation}
The last expression is an example of what was called a superaction
in
\cite{ph:actions}; there, it was supposed that
\begin{equation}
D_{\alpha (i}L_{jklm)} =0 \: \: \: \: D_{\adot (i} L_{jklm)} =0
\label{eq:Lconsts}
\end{equation}
and this is certainly sufficient for invariance of $I$ (under
Poincar\'e
supersymmetry).  However, if $L_{ijkl}$ is the first component of a
$G$-analytic field ${\cal L}$ one finds that
\begin{equation}
D_{\alpha (i}L_{jklm)} \sim D_{\alpha}^nL_{ijklmn}
\end{equation}
so that the superaction is a particular case of a harmonic integral
for an
${\cal L}$ which is CR-analytic, instead of just $G$-analytic.
Alternatively, one can
say that the constraint on $L_{ijkl}$ necessary for the superaction
to be
invariant may be weakened from (\ref{eq:Lconsts}) to the set of
conditions
which follow from $G$-analyticity of ${\cal L}$.

The $(N,p,q)$ case is the natural generalisation of the above.  If we
let
$u_{\uR}{}^i=(u_{R'}{}^i,u_{R''}{}^i)$ and $u_{\uR'}{}^i=(u_R{}^i,
u_{R''}{}^i)$ we can write the odd part of the measure in the form
\begin{equation}
A^{i_1 \ldots i_n} A^{j_1 \ldots j_n} \bar{B}_{k_1 \ldots l_m}
\bar{B}_{l_1 \ldots l_m} D_{i_1 j_1, \ldots, i_n j_n}^{k_1 l_1,
\ldots, k_m
l_m}
\end{equation}
where
\begin{eqnarray}
A^{i_1 \ldots i_n} & = & \frac{1}{n!} \varepsilon^{\uR_1 \ldots
\uR_n}
u_{\uR_1}{}^{i_1} \ldots u_{\uR_n}{}^{i_n} \nonumber \\
B^{i_1 \ldots i_m} & = & \frac{1}{m!} \varepsilon^{\uR_1' \ldots
\uR_m'}
u_{\uR_1'}{}^{i_1} \ldots u_{\uR_m'}{}^{i_m} \nonumber \\
D_{i_1 j_1, \ldots, i_n j_n}^{k_1 l_1, \ldots, k_m l_m} & = &
D_{i_1 j_1} \ldots D_{i_n j_n} \bar{D}^{k_1 l_1} \ldots \bar{D}^{k_m
l_m}
\end{eqnarray}
with $n=N-p$, $m=N-q$.  We have
\begin{eqnarray}
A^{i_1 \ldots i_n} = \varepsilon^{i_1 \ldots i_n j_1 \ldots j_p}
\bar{a}_{j_1 \ldots j_p} \nonumber \\
B^{i_1 \ldots i_m} = \varepsilon^{i_1 \ldots i_m j_1 \ldots j_q}
\bar{b}_{j_1 \ldots j_q}
\end{eqnarray}
so that, using the $\varepsilon$-tensor to dualise the indices on
the
$D$'s, we can rewrite the measure as
\begin{equation}
d^4 x \, du \,
Y_{i_1 j_1, \ldots , i_p j_p}^{k_1 l_1, \ldots ,k_q l_q}
\widetilde{D}^{i_1 j_1, \ldots ,i_p j_p}_{k_1 l_1, \ldots ,k_q l_q}
\end{equation}
where the harmonic function $Y$ is given by
\begin{equation}
Y_{i_1 j_1, \ldots , i_p j_p}^{k_1 l_1, \ldots ,k_q l_q}=
\bar{a}_{i_1 \ldots i_p} \bar{a}_{j_1 \ldots j_p} b^{k_1 \ldots k_q}
b^{l_1 \ldots l_q} \mbox{ - traces }
\end{equation}
It belongs to the following irreducible representation ($p \geq q$)
of $SU(N)$:
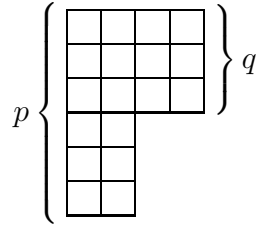
\begin{picture}(48,100)(-140,-14)
\put(0,0){\line(0,1){78}}
\put(13,0){\line(0,1){78}}
\put(26,0){\line(0,1){78}}
\put(39,39){\line(0,1){39}}
\put(52,39){\line(0,1){39}}
\put(0,0){\line(1,0){26}}
\put(0,13){\line(1,0){26}}
\put(0,26){\line(1,0){26}}
\put(0,39){\line(1,0){52}}
\put(0,52){\line(1,0){52}}
\put(0,65){\line(1,0){52}}
\put(0,78){\line(1,0){52}}
\put(-20,36)
{$p\left\{ \phantom{\begin{array}{c} . \\ . \\ . \\ . \\ .
\\ \times \end{array} } \right. $}
\put(41,56)
{$\left. \phantom{\begin{array}{c} . \\ . \\ .
\end{array} } \right\} q$}
\end{picture}

Integrating over $u$ we again pick out the first component in the
harmonic
expansion of ${\cal L}$ so that
\begin{equation}
I = \int d \mu {\cal L} = \int d^4 x
\widetilde{D}^{i_1 j_1, \ldots ,i_p j_p}_{k_1 l_1, \ldots k_q l_q}
L_{i_1 j_1, \ldots, i_p j_p}^{k_1 l_1, \ldots, k_q l_q}
\end{equation}
where
\begin{equation}
{\cal L}=\bar{Y}_{k_1 l_1, \ldots, k_q l_q}^{i_1 j_1, \ldots, i_p
j_p}
L_{i_1 j_1, \ldots, i_p j_p}^{k_1 l_1, \ldots, k_q l_q} + \ldots
\end{equation}
Again, if ${\cal L}$ is CR-analytic and not just $G$-analytic, its
harmonic expansion stops at the first term and the constraints
imposed on $L$
by analyticity are precisely those given in \cite{ph:actions}.  The
set of all
harmonic measures therefore corresponds to the set of superactions
which
have the total number of $D$'s and $\bar{D}$'s greater
than or equal to $2N$.

As an example, consider $(N,p,q)=(8,4,4)$.  In this case,
$\flag=Gr_4(8)$ so the isotropy group is $S(U(4) \times U(4))$.  The
measure has $U(1)$ charge -16, while the supergravity field strength
superfield
$W$ has charge 4.  Therefore, the harmonic integral
\begin{equation}
I= \int d \mu \, {\cal L}
\end{equation}
with
\begin{equation}
{\cal L}=W^4
\end{equation}
is manifestly Poincar\'e supersymmetric.  It is in fact the
linearised
$N=8$ supergravity three-loop counterterm.  If one integrates over
the
coset space one recovers the superaction form given in
\cite{ph:actions} with
the
Lagrangian $L$ being $W^4$ in the $4 \times 4$ square tableau
representation
of $SU(8)$.

\section{Super Yang Mills}
\label{sec:E}
The constraints on the field strength two-form in super Yang-Mills
are (with
$N\leq 4$ and a real gauge group) \cite{sohnius},
\begin{eqnarray}
F_{\alpha i \beta j} & = & \varepsilon_{\alpha \beta} W_{ij}
\nonumber \\
F_{\adot}^i{}_{\bdot}^j & = & \varepsilon_{\adot \bdot}
\bar{W}^{ij}\nonumber\\
F_{\a i}{}_{\bdot}^{j} &=& 0
\end{eqnarray}
where the field strengths are defined as usual. In $N=4$ the field
strength
$W$ is self-dual:
\beq
\bar W^{ij}={1\over2} \vare^{ijkl} W_{kl}
\eeq
The above constraints are off-shell for $N=1,2$ and on-shell for
$N=3,4$.

If we consider the same theories on $\hat M_N$, $N\geq 2$, the
constraints can

be restated in the form
\beqa
F_{\alpha R \beta S} & = & \varepsilon_{\alpha \beta} W_{RS}
\nonumber \\
F_{\adot}^{R'}{}_{\bdot}^{S'} & = & \varepsilon_{\adot \bdot}
\bar{W}^{R'S'}\nonumber\\
F_{\a R}{}_{\bdot}^{S'} &=& 0
\eeqa
where $F_{\a R\b S}=u_R{}^i u_S{}^j F_{\a i\b j}$, etc. and $R,R'$
run from 1
to
$p$ and 1 to $q$ as usual.
For $(p,q)=(1,1)$ the right-hand sides vanish and so the constraints
of $N=2,3$

Yang-Mills can be understood as
vanishing curvatures in (1,1) harmonic superspace. For $N=4$ these
constraints

still need to be supplemented by
self-duality. Hence in $N=2,3$ the constraints can be solved in
terms of pure gauges; for $N=2$ this gives the harmonic form of the
off-shell
theory \cite{gikos:N=2}, whereas in $N=3$ it gives the field
equations,
although one can then go
off-shell by allowing the field strength to be non-zero in the
$\flag$-directions \cite{gikos:N=3}.

We shall consider here two other cases, namely, $(4,2,2)$ and
$(3,2,1)$
harmonic
superspaces. In both cases we have
\beqa
F_{\a R \b S} & = & \varepsilon_{\alpha \beta} \vare_{RS} W\nonumber
\\
F_{\a R}{}_{\bdot}^{S'} &=& 0
\eeqa
For $N=3$ the third equation is
\beq
F_{\adot}^{R'}{}_{\bdot}^{S'} =0
\eeq
while for $N=4$ we have
\beq
F_{\adot}^{R'}{}_{\bdot}^{S'}  =  \varepsilon_{\adot \bdot}
\vare^{R'S'} W
\eeq
The Bianchi identities then imply that $W$ is covariantly G-analytic
\beq
\nabla_{\a R}W=\nabla_{\adot}^{R'}W=0
\eeq
and $\flag$-analytic
\beq
D_R{}^{S'}W=0
\eeq
where $\nabla$ is the Yang-Mills covariant derivative. In $N=4$ the
self-duality condition is implemented
by demanding that $W$ be real, as in the linearised case, $W=\tilde
W$.

It therefore seems to be the case that the $(1,1)$ superspace
formalism is
more powerful since
it allows one to solve the constraints. However, it is at least
interesting
that the non-linear on-shell $N=3$ or 4 theory has such a simple
description
in terms of a one-component superfield obeying simple constraints. A
small
application of this formalism is the construction of the $N=4$
supercurrent
superfield, $J$, for the fully interacting Yang-Mills theory. It is
given by
\beq
J={\rm Tr} \, (W^2)
\eeq
and is clearly CR-analytic and real.
It has $U(1)$ charge 4 and couples to a linearised
potential for $N=4$ conformal supergravity, $V$, by
\beq
\int d\m \: J \, V
\eeq
where $d\m$ is the (4,2,2) measure discussed in the previous section.

The potential $V$ is G-analytic, real and has charge 4. The
gauge transformations which leave the above interaction invariant
are
\beq
\d V= D_R{}^{S'} X_{S'}{}^R
\eeq
where the gauge parameter $X$ is $G$-analytic and real.

\section{Conformal Supergravity}
\label{sec:F}
In this section we shall apply the harmonic superspace formalism to
supergravity theories, in particular, conformal supergravity theories
which
exist for $N=1,2,3,4$. The superspace constraints for these theories
were
written down in \cite{ph:csg}. For $N=2$ our results give a
geometrical
interpretation
of the work of \cite{gikos:N=2sg,gikos:N=2sg2} on the subject, while
for
$N=3,4$ we show how this
geometry generalises to higher $N$, although we shall not go into any

detail about solving the constraints here. In fact the same formalism
can
also be applied to $N=5,6,7,8$, and in these cases we find that the
constraints

imposed by harmonic superspace considerations lead to the conformal
constraints proposed in \cite{ph:sg}; these contraints are not fully
off-shell,
as was
noted in \cite{gg}, but they are compatible with the field equations
of
on-shell
Poincar\'e supergravity \cite{bh,ph:sg}. Thus the geometry proposed
below
summarises a large
part of the known results on superspace supergravity from the
harmonic point
of view.

In order to derive these results we shall have to make some
assumptions about
the basic geometrical structures that are to be imposed. We start
with a
$(4|4N)$-dimensional real supermanifold $M$ which is equipped with a
choice
of odd tangent bundle, $F$ (rank $(0|4N)$), $F\subset T$, where $T$
is the
tangent bundle. In addition we shall suppose that $F$ is maximally
non-integrable in the sense that
\beq
[F,F] \mbox{ mod } F=B:= T/F
\eeq
that is, the even tangent bundle, $B$, defined as the quotient of $T$
by $F$
is spanned by the commutators of odd vector fields. We shall further
suppose
that the structure group of the bundle of odd frames $LF$ can be
reduced from
$GL(4N,\real)$ to $SL(2,\com)\cdot U(N)\xz \real^+$ where
$\real^+$ corresponds to Weyl rescalings. This implies (at least
locally) that

the complexification of $F$, $F_c$ can be written as
${\cal F}\oplus \bar{\cal F}$ where ${\cal F}=S\otimes V$, with $S$
having
rank $(0|2)$ and $V$ having rank $(N|0)$. The above structure defines
a tensor
$T$ which we shall call the structure tensor of the theory and which
is a
section of $\wedge^2 F^*\otimes B$.
If we introduce local basis
vector fields $\{E_{\a i},\bar E_{\adot}^i, E^a\}$ for ${\cal F},\bar
{\cal F}$

and $B^*$ respectively, the components of the structure tensor can
be expressed as
\beqa
T_{\a i\b j}{}^c &=& -<[E_{\a i},E_{\b j}], E^c> \nonumber \\
T_{\adot}^i{}_{\bdot}^{jc} &=& -<[E_{\adot}^i,E_{\bdot}^j], E^c>
\nonumber \\
T_{\a i}{}_{\bdot}^{jc} &=& -<[E_{\a i},E_{\bdot}^j], E^c>
\label{96}
\eeqa
the second equation being the complex conjugate of the first.

The conformal constraints are simply that the components of $T$ are
the same
as in the flat case, i.e.
\beq
T_{\a i\b j}{}^c=T_{\adot}^i{}_{\bdot}^j{}^c=0
\eeq
and
\beq
T_{\a i}{}_{\bdot}^j=-i(\s^c)_{\a\bdot}\d_i{}^j
\eeq
More precisely, if $T$ is flat, one can choose a basis for $B$ as a
subbundle
of $T$ and an ${\gsl}(2)\oplus {\gu}(N)$ connection such that the
conformal
constraints are recovered.

The result which we shall demonstrate here is that the above
constraints can
be interpreted geometrically as the statement that $(1,1)$ harmonic
super\-space is a CR supermanifold with CR bundle $K$ of rank
$(2N-3|4)$. In
the gravitational context harmonic superspace is the $U(N)$ bundle
with fibre
$\flag$(=$\flag_{p,N-q}(N)$ in the $(p,q)$ case) associated with the
principal

$U(N)$ bundle (assuming it exists) which corresponds to the $U(N)$
part of
the structure group.  It is often convenient to work with $U(N)$
instead of
$SU(N)$ in supergravity and $\flag$ can be
viewed as a coset space of $U(N)$ with isotropy group
$U(p)\xz U(q)\xz U(N-(p+q))$.
If we denote the principal $U(N)$ bundle by $\hat M_N$ we can
study harmonic superspace $M_N(p,q)$ by working with fields on $\hat
M_N$
that are equivariant with respect to the isotropy subgroup.

To prove the result we first split $T=F\oplus B$ and introduce an
$\gsl(2,\com)\oplus \gu(N)$ connection $\C$. If
$E_A=\{E_{\a i},\bar E_{\adot}^i,E_a\}$ denotes a set of local basis
vectors
for $T$, the horizontal lifts, $\hat E_A$, of these vector fields in
$\hat M_N$

are defined by
\beq
\hat E_A=E_A-\C_{AI}{}^J D_J{}^I
\eeq
where $D_I{}^J$ are now the right-invariant vector fields on $U(N)$
defined by
\beq
D_I{}^J=u_I{}^i{\partial\over\partial u_J{}^i}
\eeq
and
\beq
\C_{AI}{}^J= u_I{}^i\C_{Ai}{}^j u_j{}^J
\eeq
is the $\gu(N)$ part of the connection.
A local basis for $\hat M_N$ is given by $\{\hat E_A, D_I{}^J\}$ The
claim is
that the basis vectors of the CR subbundle, $K$, of the tangent
bundle of
$M_N(1,1)$ are the following:
\beq
D_1{}^r, D_r{}^N, D_1{}^N;\  \hat E_{\a 1},\hat{\bar E}_{\adot}^N
\eeq
where $r=2,\dots (N-1)$ and
\beq
\hat E_{\a I}=u_I{}^i \hat E_{\a i};\qquad \hat{\bar
E}_{\adot}^I=u_i{}^I
\hat{\bar E}_{\adot}^i
\eeq
We have
\beq
[\hat E_{\a I},\hat E_{\b J}]=
-T_{\a I\b J}{}^C \hat E_C + \C_{\a I,\b}{}^\c
\hat E_{\c J}+\C_{\b J,\a}{}^\c \hat E_{\c I}- R_{\a I,\b J,K}{}^L
D_L{}^K
\eeq
and similarly for dotted indices where $\C_{A, \beta}{}^{\gamma}$ is
the
$\gsl(2,\com)$ part of the connection, the little $U(N)$ indices are

converted to capital ones by using $u$ as usual, and $R$ is the
$U(N)$
curvature tensor. Since we are interested in establishing the
involutivity of
$K$ in $M_N(1,1)$ we can take the above vector fields to be act on
functions
which are invariant under the isotropy group $U(1)\xz U(N-2)\xz
U(1)$. The
conditions for involutivity of $K$ therefore include, at dimension
zero,
\beqa
T_{\a 1\b 1}{}^c &=& T_{\adot}^N{}_{\bdot}^{N c}=0 \nonumber \\
T_{\a 1}{}_{\bdot}^{Nc} &=& 0
\eeqa
at dimension one-half,
\beqa
T_{\a 1\b 1,}{}_1^{\gdot}&=&T_{\a 1\b 1,}{}_r^{\gdot}=0\nonumber\\
T_{\adot}^N{}_{\bdot}^{N\c N} &=&T_{\adot}^N{}_{\bdot}^{N\c r}=0
\eeqa
and at dimension one,
\beq
R_{..,1}{}^r=R_{..,r}{}^N= R_{..,1}{}^N=0
\eeq
where the missing indices can be any of the pairs
$\{{}_{(\a 1,\b1)}, ({}_{\a 1},{}_{\bdot}^N),
({}_{\adot}^N,{}_{\bdot}^N)\}$,
and where $T$ now denotes the torsion tensor. The
dimension zero
constraints imply flatness of the structure tensor (which becomes the
dimension
zero part of the
torsion tensor) when supplemented by algebraic constraints, and the
rest can
be shown to be consistent after using a little algebra.

The use of the group $U(N)$ is particularly useful in conformal
supergravity
since the corresponding spacetime gauge field belongs to the
conformal
supermultiplet. However, for $N=4$, there are only gauge fields for
the $SU(4)$
subgroup, and so an additional constraint is needed in this case
\cite{ph:csg}.

The computation carried out here can be repeated for any choice of
$(p,q)$.
It is found that, for $N=3$, $(3,2,1)$ harmonic superspace is a CR
supermanifold, and for $N=4$ $(4,2,2)$ superspace is a CR
supermanifold.
However, if one lifts the structure tensor up to $M_N(p,q)$ by
replacing the
$E$'s by the $\hat E$'s in (\ref{96}), one sees that the connections
drop out.
Hence
the constraints can also be generated by demanding that the lifted
structure
tensor vanish along $K$ for any choice of $(p,q)$ even though $K$ may
not be
involutive.

The geometry of $N=1$ conformal supergravity does not fit into the
above scheme
due to the fact that the internal symmetry group is in this case only
$U(1)$.
However, the Ogievetsky-Sokatchev formalism \cite{os,siegel} for
complexified
supergravity can be described by the double fibration
$\mink_L\leftarrow
\mink\rightarrow\mink_R$, where $\mink_L(\mink_R)$ are respectively
the left
and right chiral superspaces \cite{manin}. In the real case the
intrinsic
geometrical formulation of $N=1$ conformal supergravity may be stated
as
follows. $M$ is a real $(4|4)$  CR supermanifold with CR bundle
${\cal F}$ of
rank $(0|2)$ such that
\beq
[{\cal F},\bar{\cal F}]\, {\rm mod}\, F_c =B_c
\eeq
where $F_c={\cal F}\oplus\bar{\cal F}$ and $B_c=T_c/F_c$. Given this
structure
one can reconstruct the OS formalism straightforwardly.

In conclusion, we have seen that the harmonic superspace formalism,
for various
values of $(p,q)$, can be used to describe the constraints of
$N=2,3,4$
conformal supergravity and the conformal constraints of $N\geq 5$
supergravity.
In the conformal case one would like to solve these constraints to
obtain new
potentials for the $N=3,4$ theories. However, this is not as simple
as in
$N=2$, since the complex dimension of the internal flag manifold is
greater
than one. In addition, one can see on dimensional grounds that there
will be
complications compared to $N=2$; for example, the linearised harmonic
potential
$V$ for $N=4$ introduced in the previous section has dimension $-2$.
In the
case of Poincar\'e supergravity one needs to find the geometrical
formulation
of the additional constraints that arise; it is possible that
extended
supergravity theories with sufficiently large $N$ may turn out to be
in some
sense integrable, that is, the constraints defining the field
equations are
given by vanishing curvatures on some appropriate superspace.

\end{document}